\documentclass[iop,apj,twocolumn]{emulateapj_pab}	

\RequirePackage{hyperref}
\hypersetup{
	colorlinks = true,
	linkcolor = blue,
	citecolor = blue,
	filecolor = blue,
	urlcolor = blue,
	pdfborder = {0 0 0}
}

\usepackage{color,bm,mathrsfs,amsmath}

\renewcommand{\vec}[1]{\bm{#1}}

\graphicspath{{figures/}{../figures/}}

\submitted{Received 2018 April 11; revised 2018 October 17; accepted 2018 October 17; published 2018 December 5}
\journalinfo{The Astrophysical Journal, {\rm 869:2 (9pp), 2018 December 10 \hfill \url{https://doi.org/10.3847/1538-4357/aae97a}}}
\shorttitle{Magnetic helicity reversal in the corona}
\shortauthors{Bourdin, Singh, \& Brandenburg}

\newcommand{\sect}[1]{Section\,\ref{S:#1}}

\DeclareRobustCommand*{\fig}[1]{Figure\,\ref{F:#1}}

\newcommand{\eqn}[1]{Equation\,(\ref{E:#1})}

\newcommand{\graphflex}[4][figure]{\begin{#1}#2\caption{#4\label{F:#3}}\end{#1}}
\newcommand{\graph}[3]{\begin{figure}\plotone{#1.pdf}\caption{#3\label{F:#2}}\end{figure}}
\newcommand{\graphwidthflex}[6][figure*]{\graphflex[#1]{#5\includegraphics[width=#4]{#2.pdf}}{#3}{#6}}
\newcommand{\graphwidth}[4][15cm]{\graphwidthflex{#2}{#3}{#1}{\centering}{#4}}
\newcommand{\graphfull}[3]{\graphwidth[17cm]{#1}{#2}{#3}}
\newcommand{\graphside}[3]{\graphwidth[13.2cm]{#1}{#2}{#3}}
\newcommand{\eql}[1]{\begin{equation}#1\end{equation}}
\newcommand{\eqa}[1]{\begin{eqnarray}#1\end{eqnarray}}

\newcommand{\eqi}[1]{$#1$}

\def\half{{\textstyle{1\over2}}}
\newcommand{\m}{\,{\rm m}}
\newcommand{\cm}{\,{\rm cm}}
\newcommand{\Mm}{\,{\rm Mm}}
\newcommand{\Mx}{\,{\rm Mx}}
\newcommand{\s}{\,{\rm s}}

\newcommand{\braxy}[1]{\langle #1\rangle_{xy}}
\newcommand{\SSSS}{\mbox{\boldmath ${\sf S}$} {}}

\DeclareRobustCommand*{\unit}[1]{\def~{\,}\ensuremath{\mathrm{\,#1}}}
\definecolor{darkgreen}{rgb}{0,0.45,0}

\begin{document}
\hypersetup{
	pdftitle = {Magnetic Helicity Reversal in the Corona at Small Plasma Beta},
	pdfauthor = {Philippe-A.~Bourdin},
	pdfkeywords = {Sun: corona -- Sun: magnetic fields -- solar wind -- dynamo -- magnetohydrodynamics (MHD) -- methods: numerical},
	pdfsubject = {The Astrophysical Journal, 869(2018) 2. doi:10.3847/1538-4357/aae97a}
}


\title{Magnetic Helicity Reversal in the Corona at Small Plasma Beta}

\author{Philippe-A.~Bourdin$^{1}$, \orcid{0000-0002-6793-601X},
Nishant~K.~Singh$^{2,3}$, \orcid{0000-0001-6097-688X},
Axel~Brandenburg$^{3,4,5,6}$, \orcid{0000-0002-7304-021X}
}
\email{Philippe.Bourdin@oeaw.ac.at}

\affiliation{$^{1}$ Space Research Institute, Austrian Academy of Sciences, Schmiedlstr. 6, A-8042 Graz, Austria}
\affiliation{$^{2}$ Max-Planck-Institut f{\"u}r Sonnensystemforschung, Justus-von-Liebig-Weg 3, D-37077 G{\"o}ttingen, Germany}
\affiliation{$^{3}$ Nordita, KTH Royal Institute of Technology and Stockholm University, Roslagstullsbacken 23, SE-10691 Stockholm, Sweden}
\affiliation{$^{4}$ JILA and Department of Astrophysical and Planetary Sciences, University of Colorado, Boulder, CO 80303, USA}
\affiliation{$^{5}$ Department of Astronomy, AlbaNova University Center, Stockholm University, SE-10691 Stockholm, Sweden}
\affiliation{$^{6}$ Laboratory for Atmospheric and Space Physics, University of Colorado, Boulder, CO 80303, USA}

\begin{abstract}
Solar and stellar dynamos shed small-scale and large-scale magnetic
helicity of opposite signs.
However, solar wind observations and simulations have shown that
some distance above the dynamo both the small-scale and large-scale
magnetic  helicities have reversed signs.
With realistic simulations of the solar corona above an active region
now being available, we have access to the magnetic field and current density
along coronal loops.
We show that a sign reversal in the horizontal averages of the
magnetic helicity occurs when the local maximum of the plasma beta
drops below unity and the field becomes nearly fully force free.
Hence, this reversal is expected to occur well within the solar corona
and would not directly be accessible to in-situ measurements with
the {\em Parker Solar Probe} or {\em SolarOrbiter}.
We also show that the reversal is associated with subtle changes in the
relative dominance of structures with positive and negative magnetic helicity.
\end{abstract}
\keywords{ Sun: corona --- Sun: magnetic fields --- solar wind --- dynamo --- magnetohydrodynamics (MHD) --- methods: numerical }

\section{Introduction} \label{S:helicity.intro}

Magnetic helicity is an important invariant in ideal and nearly ideal
magnetohydrodynamics (MHD); see \cite{Biskamp:2003}
and the original work of \cite{Woltjer:1958}.
It plays a crucial role in characterizing the topological complexity
of coronal magnetic fields \citep{Berger+Field:1984}, and it is also
responsible for the possibility of premature quenching of the underlying
dynamo \citep{Gruzinov+Diamond:1994}.
An obvious remedy to the dynamo problem is to let excess magnetic helicity
escape through the boundaries \citep{Blackman+Field:2000,Kleeorin+al:2000},
especially through coronal mass ejections \citep{Blackman+Brandenburg:2003}
and the solar differential rotation, which also acts on open field lines
rooted in magnetically quiet regions of the photosphere \citep{DeVore:2000};
see \cite{Berger+Ruzmaikin:2000} for estimates of the relative importance
of different contributions to the total magnetic helicity flux.
Much of the magnetic helicity transported by differential rotation out through
the surface has entered through the equator; see \cite{Brandenburg+Sandin:2004}.
Its contribution to the net magnetic helicity loss may therefore be subdominant.
Nevertheless, estimates for the Sun invariably result in a total loss
of $\mp10^{46}\Mx^2$ per 11 year cycle \citep{Berger+Ruzmaikin:2000,
DeVore:2000, Brandenburg+Sandin:2004, Brandenburg:2009} in the northern
and southern hemispheres, respectively.
One would therefore expect to see that the magnetic helicity
shed at the solar surface agrees with what is passing through the
solar wind at larger distances.
However, this does not seem to be the case because in the solar wind,
the magnetic helicity was found to have mostly a positive sign in the
northern heliosphere \citep{Brandenburg+al:2011}, whereas at the solar
surface it is mostly negative in the north \citep{Seehafer:1990}.
A similar result was also obtained in numerical simulations of dynamos
with a coronal exterior \citep{Warnecke+al:2011,Warnecke+al:2012}.
This unexpected behavior is what is referred to as magnetic
helicity reversal.
Such reversals have also been found in analytic solutions of simple dynamo
models with a force-free exterior \citep{Bonanno:2016} and in mean-field
models with magnetic helicity fluxes included \citep{BCC09}.

Numerical simulations are currently the best way of testing and studying
in detail the idea of the magnetic helicity reversal.
Here, we consider the magnetic helicity that was found to emerge from
the MHD simulations of \citet*[][hereafter BBP]{Bourdin+al:2013_overview},
who used a solar magnetogram of an active region (AR) as a boundary
condition.
Reconnection and the associated coronal heating were driven by random
footpoint motions, as envisaged in the early work of \cite{Parker:1972}.
The simulations of BBP used the
{\sc Pencil Code}\footnote{\url{https://github.com/pencil-code}}
and covered a larger domain ($235^2\times156\Mm^3$) compared to earlier
ones with the {\sc Stagger Code} \citep{Gudiksen+Nordlund:2002,
Gudiksen+Nordlund:2005a,Gudiksen+Nordlund:2005b}.

The AR model of BBP is observationally driven by
line-of-sight magnetograms taken from {\sc Hinode/SOT-NFI}
\citep{Kosugi+al:2007,Tsuneta+al:2008}.
The model provides a sufficient amount of energy to the corona \citep{Bourdin+al:2015_energy-input}.
It also compares well with various coronal observations (BBP) and
shows similarities to coronal scaling laws, e.g., for the temperature
along loops that were derived from earlier observational and theoretical
works \citep[cf.][]{Bourdin+al:2016_scaling-laws}.

The coronal EUV emission is synthesized from the MHD model using the
{\sc Chianti} atomic database \citep{Dere+al:1997,Young+al:2003}
using the method of \cite{Peter+al:2004,Peter+al:2006}.
The 3D structure of the AR loop system matches the
reconstruction from {\sc Stereo} observations.
Also, the plasma flow dynamics along those loops matches the Doppler
shift pattern observed by {\sc Hinode/EIS} \citep{Culhane+al:2007}
in the coronal \ion{Fe}{12} emission line.

The magnetograms for driving these simulations from the bottom boundary give
just the line-of-sight magnetic field, or $B_z$ near disk center.
During the first hour of solar time, we do not yet apply any large-scale
driving motions derived from the observed movements of magnetic patches in
the photosphere.
We only apply the horizontal small-scale velocities that mimic granulation.
Therefore, these photospheric horizontal motions are purely stochastic
and statistically mirrorsymmetric, and there is no obvious mechanism
to break the statistical mirrorsymmetry of the model.
In particular, there is no Coriolis force or differential rotation.
Nonetheless, it turns out that helicity emerges readily within the
initial phase of our model.

Although there is no direct injection of helicity, the model can still
produce magnetic helicity through a complex arrangement of multipolar
spots \citep{Bourdin+Brandenburg:2018_multipole}.
We use a magnetogram of a small and stable AR observed
during 2007~November~14 in the southern hemisphere.
Indeed, we find helicity, as is readily demonstrated
by looking at the vertical profile of the mean current
helicity density, $\braxy{\vec{J}\cdot\vec{B}}$, where
$\vec{J}=\vec{\nabla}\times\vec{B}/\mu_0$ is the current density,
$\vec{B}$ is the magnetic field, $\mu_0$ is the vacuum permeability,
and $\braxy{...}$ denotes horizontal averages.
We find that the profiles generally show a sign reversal within the first
$5$--$15\unit{Mm}$ above the surface.
This is equally remarkable because the upper regions are topologically
connected with the lower ones through the same large-scale structures.
Any small-scale magnetic fields seem to be interspersed within other
structures and are still associated with the large-scale magnetic loops
extending from one footpoint to the other.

The purpose of this work is to quantify the magnetic helicity reversal in
detail and to associate it with coronal heating along EUV-emissive loops.
We further characterize the magnetic helicity reversal in spectral space
and demonstrate that it occurs in different wavenumber intervals at the
same height.

\section{Our approach}

In the present work, we use a snapshot from the simulations of BBP to
analyze the production and vertical variation of magnetic and current
helicity densities as well as their spectra.
Before discussing those aspects in detail, we begin with the basic
equations solved in BBP and present a brief summary of the physical
properties of those simulations.

\subsection{Basic equations}\label{S:Basic.Eqns}

BBP solved the continuity equation, the equation of motion, the induction
equation, and an energy equation, which includes the necessary energy sinks
to get realistic and self-consistent coronal heating and cooling terms:
\eqa{\frac{{\rm{D}}\ln \rho}{{\rm{D}}t} &=& -\vec{\nabla} \cdot \vec{u}, \label{E:continuity} \\
\rho\frac{{\rm{D}}\vec{u}}{{\rm{D}}t} &=& -\vec{\nabla}P + \rho\vec{g}
+ \vec{J} \times \vec{B} + \vec{\nabla}\cdot(2\nu\rho\SSSS),
\label{E:motion} \\
\rho\,T\,\frac{{\rm{D}}s}{{\rm{D}}t} &=&
- \vec{\nabla} \cdot \vec{F} - \rho^2\Lambda(T) +\mu_0\eta\vec{J}^2 + 2\rho\nu{\SSSS}^2, \label{E:energy}\\
\frac{\partial\vec{A}}{\partial{t}} &=& \vec{u} \times \vec{B} - \mu_0\eta\vec{J}, \label{E:induction}
}
where $P=({\cal R}/\mu)\,\rho T$ is the gas pressure,
${\cal R}$ is the universal gas constant, $\mu=0.67$ is the
mean atomic mass, $T$ is the temperature,
$F_i=-\rho c_P \chi_{ij}\nabla_j T$ is the conductive heat flux,
$s=c_V\ln P-c_P\ln\rho+s_0$ is the specific entropy, $s_0$ is a constant,
\eqi{c_P} and \eqi{c_V} are the specific heats at constant pressure
and constant volume, respectively,
$\chi_{ij}=\chi_0\delta_{ij}+\chi_{\rm Spitz}\hat{B}_i\hat{B}_j$
is the thermal diffusivity, $\vec{\hat{B}}$ is the unit vector
of the magnetic field, \eqi{\nu} is the kinematic viscosity,
\eqi{\eta} is the magnetic diffusivity,
\eqi{\chi_{\rm Spitz}} is the Spitzer field-aligned heat conductivity,
\eqi{\chi_0} is an isotropic contribution,
$\vec{g}$ is the gravitational acceleration, and
$\SSSS$ is the traceless rate-of-strain tensor with the components
$\mathsf{S}_{ij}=\frac{1}{2}(u_{i,j}+u_{j,i})-\frac{1}{3}\delta_{ij}u_{k,k}$.
Here, we solve for $\vec{A}$ because then $\vec{B}$ is automatically
divergence free.
Instead of solving for $s$, we use the logarithmic temperature $\ln T$,
which is directly related to $s$ and $\ln\rho$.
Using the logarithmic density $\ln\rho$, we are able to capture
many orders of magnitude in the density stratification that our model
atmosphere covers.

\subsection{Physical details about the simulations}\label{S:sim.details}

For the radiative cooling function $\rho\Lambda(T)$, we use
a realistic tabulation of $\Lambda$ provided by \cite{Cook+al:1989}.
In that work, the important emission peak from highly ionized iron
lines is included, which efficiently cools the model corona.
The characteristic half-time of this cooling is below 30 minutes.

During the first 35 minutes of physical time, however, this loss term
is turned off together with the heat conduction along the field.
This is to prevent excessive cooling of the corona during the
initial phase in which the granular motions cause magnetic disturbances that
still need to propagate from the photosphere into the corona.
The simulation is then continued for another about 35 minutes after all
physical terms in the equations are turned on.
Further details about the switching on can be found in
\cite{Bourdin+al:2014_coronal-loops} and \cite{Bourdin:2014_switch-on}.
We use here data from a fully developed state at 63 minutes physical time.

Unlike ideal models of MHD, where $\eta=\nu=0$ and dissipation is
modeled by highly nonlinear diffusion operators that cannot easily be
stated in concise form and effective Reynolds or Lundquist numbers are
difficult to specify, we use in our model constant values of $\nu$,
$\eta$, and also $\chi_0$.
The value of the magnetic diffusivity $\eta=10^{10}\m^2\s^{-1}$ ($=10^{14}\cm^2\s^{-1}$)
is about eight orders of magnitude bigger than that estimated for the solar corona.
This choice is required for numerical stability and for having a grid
Reynolds number near unity.
On the other hand, we use a realistic value for the viscosity, $\nu=10^{10}\m^2\s^{-1}$,
which results in a Prandtl number of unity because $\eta=\nu$.
However, as pointed out by \cite{Rempel:2017}, the relative importance of
Ohmic and viscous heating changes toward the latter when realistically
large values of the magnetic Prandtl numbers are taken into account;
see \cite{Brandenburg:2014} for the relation between the dissipation
ratio and the magnetic Prandtl number.
The isotropic heat conduction is set to $\chi_0=5\times10^8\m^2\s^{-1}$
($=5\times10^{12}\cm^2\s^{-1}$).
We use a realistic coronal value of
$\kappa_{\rm Spitz}=1.8 \cdot 10^{-10}\,T^{5/2}/\ln \Lambda_{C}$
with \eqi{\ln \Lambda_{C} = 20} being the Coulomb logarithm.
We obtain the field-aligned Spitzer conductivity as
\eql{\chi_{\rm Spitz} = \frac{1}{c_P} \kappa_{\rm Spitz} T^{5/2} / \rho .}

\subsection{Boundary conditions}

The model is periodic in the horizontal directions and employs a
potential-field extrapolation on the top boundary.
To formulate the potential-field boundary condition, we define
the Fourier-transformed magnetic vector potential as
\eql{\tilde{\vec{A}}(k_x,k_y,z,t)=\int\vec{A}(x,y,z,t)\,
e^{{\rm{i}}\vec{k}\cdot\vec{r}} {\rm{d}}^2\vec{r},}
where $\vec{k}=(k_x,k_y)$ and $\vec{r}=(x,y)$.
On the lower and upper $z$ boundaries, we thus have
\eql{{\partial\tilde{\vec{A}}\over\partial z}=-|\vec{k}|\tilde{\vec{A}}(k_x,k_y,z_\ast,t),
\label{E:potBC}}
where $z_\ast$ denotes the locations of the boundaries.
Apart from this, the top boundary is closed for any plasma flows and
is thermally insulating.

At the bottom boundary, realistic atmospheric temperature and density are imposed.
We also adopt \eqn{potBC} at the bottom boundary,
but keep the minus sign on the right-hand side.
This corresponds to an inverted potential-field extrapolation, whereby
contrasts in the magnetic field are increased, instead of smearing them
out like for the top boundary.
This mimics the effect of flux tubes becoming narrower when entering below the photosphere.
Because this increase of contrast would quickly lead to artifacts, like
wiggles in the $B_z$ component, this increase of contrast is limited
to about one-third of the pressure scale height some 100\unit{km}
below the photosphere so that artifacts in $B_z$ are avoided.
With this method, we obtain the ghost zones for all three components of
$\vec{A}$ just beneath the lower photospheric boundary.

\subsection{Gauge dependence of magnetic helicity} \label{S:gauge.dependence}

In general, the local magnetic helicity density
\eql{{\cal H}_{\rm M} = \vec{A} \cdot \vec{B} \label{H_M}}
depends on the gauge of the vector potential \eqi{\vec{A}}.
On the lower and upper boundaries of the simulation domain, our
magnetic field is driven toward a potential state. The resulting
\eqi{\vec{A}} at these boundaries is in the Weyl gauge.
However, the gauge could in principle still drift because we have no boundary
restrictions other than periodicity along \eqi{x} and \eqi{y}.
Therefore, we must check if such a gauge drift occurs and if it
significantly changes our simulation results.

We use the relative helicity from Equation~(5)
of \cite{Finn+Antonsen:1985}, similar to the formulation of
\cite{Berger+Field:1984}, to obtain a gauge-independent helicity as
\eql{{\cal H}_{\rm M,rel}(z) = \iiint_z^{\infty} {1 \over 2} (\vec{A} + \vec{A_{\rm pot}}) \cdot (\vec{B} - \vec{B_{\rm pot}})\,{\rm d}z\,{\rm d}y\,{\rm d}x \label{E:H_rel}}
\eqi{\vec{A_{\rm pot}}} and \eqi{\vec{B_{\rm pot}}} are the nonhelical potential fields extrapolated from the known state of \eqi{B_z(z)} at the height \eqi{z}.
On the upper boundary \eqi{z=L_z}, our magnetic field is already
almost potential and very close to the nonhelical extrapolation
\eqi{\vec{B_{\rm pot}}(L_z)}.
The magnetic helicity therefore vanishes toward the top of the
domain and we may omit the volume from \eqi{L_z} to \eqi{\infty} in the
integrals of \eqi{H_{\rm rel}}.
Our factor of one-half compensates for the addition of the two similar quantities \eqi{\vec{A}} and \eqi{\vec{A_{\rm pot}}} in \eqn{H_rel}, which allows the relative magnetic helicity to be quantitatively similar to the magnetic helicity \eqi{{\cal H}_{\rm M}}.
Because the components of \eqi{\vec{B}} and \eqi{\vec{{B_{\rm pot}}}}
are normal to all nonperiodic boundaries (here, the top and bottom
of the simulation domain), they are either identical by construction or their
differences are negligible. \eqn{H_rel} gives us therefore a gauge-independent
relative helicity \citep{Berger+Field:1984,Finn+Antonsen:1985}.

To compare a vertical profile of \eqi{{\cal H}_{\rm M}} and \eqi{{\cal H}_{\rm M,rel}},
we compute the horizontal averages of both quantities.
To get \eqi{\langle{\cal H}_{\rm M}(z)\rangle_{xy}} of the magnetic helicity
density \eqi{{\cal H}_{\rm M}(z)}, we simply average over horizontal slices
from the height \eqi{z} and with a thickness of \eqi{\Delta z} equal to
our vertical grid spacing.
For the profile of \eqi{{\cal H}_{\rm M,rel}(z)}, we simply subtract the
volume integrals above the heights \eqi{z} and \eqi{z+\Delta z}.
Each integral uses different potential fields,
\eqi{A,B_{\rm pot,z}} and \eqi{A,B_{\rm pot,z+\Delta z}}, which we extrapolate
from the known states \eqi{B_z(z)} and \eqi{B_z(z+\Delta z)}, respectively.
The average relative magnetic helicity density contained in this \eqi{xy} layer
is then just the difference
\eql{\langle{\cal H}_{\rm M,rel}\rangle_{xy}(z) = {{\cal H}_{\rm M,rel}(z) - {\cal H}_{\rm M,rel}(z+\Delta z) \over V_{xy\Delta z}},}
which we normalize to the volume of the layer \eqi{V_{xy\Delta z}} to be comparable to \eqi{\langle{\cal H}_{\rm M}(z)\rangle_{xy}}; see \fig{Gauge_Dependence}.

We find that the horizontal averages of our magnetic helicity density \eqi{{\cal H}_{\rm M}} and the relative helicity density \eqi{{\cal H}_{\rm M,rel}} are very similar.
Both magnetic helicities do show sign reversals that are located roughly at the same height, like the maxima and minima; see the dotted and red lines in \fig{Gauge_Dependence}.
The vertical profile of the current helicity,
\eqi{{\cal H}_{\rm C} = \mu_0 \vec{j} \cdot \vec{B}},
shows a qualitatively similar trend also with a sign reversal
in the corona, albeit higher up; see the blue dashed line.
We show that our magnetic helicity density is therefore not significantly
influenced by the gauge drift along the periodic directions.
Therefore, we may continue to use the gauge-dependent magnetic helicity
\eqi{{\cal H}_{\rm M}} as a good proxy of the gauge-independent relative
helicity \eqi{{\cal H}_{\rm M,rel}}.

\subsection{Magnetic and current helicity spectra} \label{S:helicity.spectra}

\graph{AR_helicity-height_VAR378}{Gauge_Dependence}{Horizontal averages of the gauge-dependent magnetic helicity, the gauge-independent relative helicity, and the current helicity normalized to $\vec{B}^2$ versus height.}

\graphfull{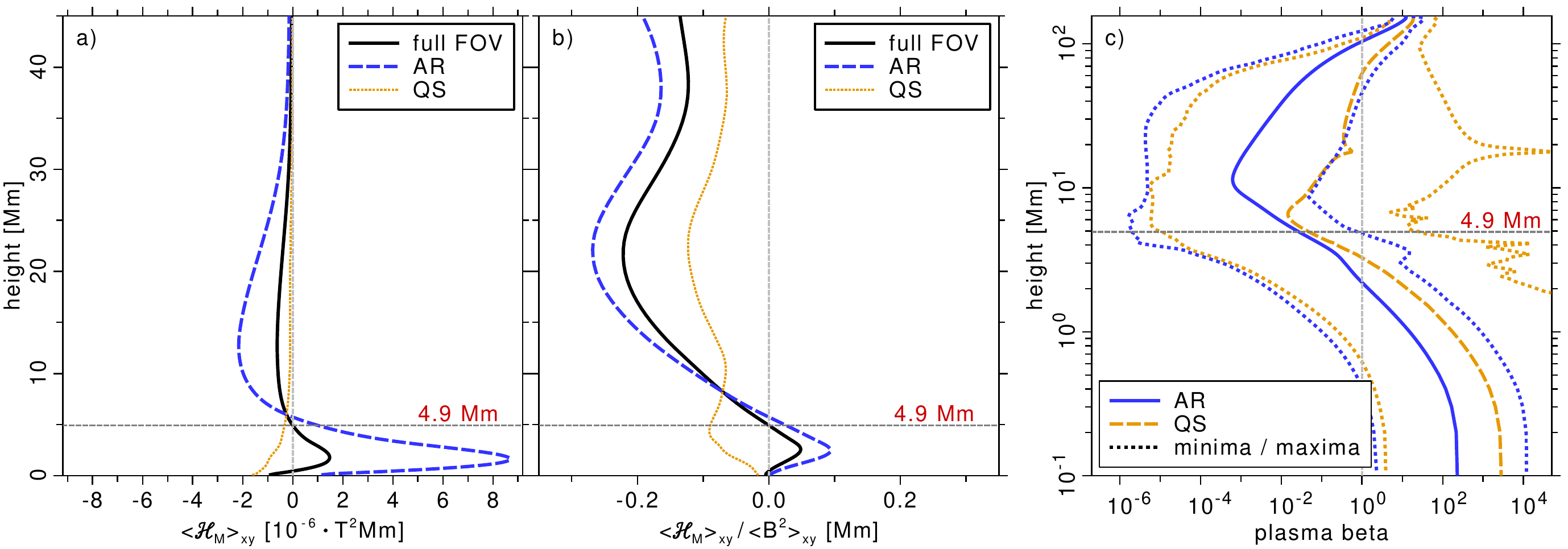}{Helicity_Height}{(a) Magnetic helicity for the full FOV, the AR core, and only the QS area as a profile of horizontal averages versus height. (b) Magnetic helicity normalized to $\vec{B}^2$. (c) Plasma beta value ranges for AR and QS.}

Through most of this work, we show both current and magnetic helicity.
In particular, we consider two-dimensional current and magnetic helicity
spectra defined as
\eqa{H_{\rm C}(k)&=\;\half\!\!\!\!\!\!\!\!&\sum_{k_- < |{\bf k}|\leq k_+}\!\!\!\!\!\!\left(
\tilde{\vec{J}}\cdot\tilde{\vec{B}}^\ast+\tilde{\vec{J}}^\ast\cdot\tilde{\vec{B}}\right),\\
H_{\rm M}(k)&=\;\half\!\!\!\!\!\!\!\!&\sum_{k_- < |{\bf k}|\leq k_+}\!\!\!\!\!\!\left(
\tilde{\vec{A}}\cdot\tilde{\vec{B}}^\ast+\tilde{\vec{A}}^\ast\cdot\tilde{\vec{B}}\right),\label{EHvec}}
where $k_\pm=k\pm\delta k/2$ and $\delta k=2\pi/L$ with $L=235\Mm$ being the
size of the magnetograms and tildes denote, again, Fourier transformation.
Under horizontally isotropic conditions, we have
\eql{\mu_0 H_{\rm C}(k)=k^2H_{\rm M}(k),\label{E:HC_k2HM}}
i.e., the current helicity spectrum is
directly related to the magnetic helicity spectrum, but weighted with a
$k^2$ factor, so high wavenumber contributions in $H_{\rm M}(k)$ get enhanced.

It is sometimes convenient to define the magnetic and current
helicity densities as
\eql{{\cal H}_{\rm M}=\vec{A}\cdot\vec{B},\quad
{\cal H}_{\rm C}=\vec{J}\cdot\vec{B}.}
Note, in particular, that at each value of $z$, we have
\eql{\int H_{\rm M}\,{{\rm d}}k=\braxy{{\cal H}_{\rm M}},\quad
\int H_{\rm C}\,{{\rm d}}k=\braxy{{\cal H}_{\rm C}}.}
In the following, however, we often retain the more explicit notation
in terms of $\vec{A}\cdot\vec{B}$ and $\vec{J}\cdot\vec{B}$.

\section{Results}

\subsection{Magnetic helicity reversal}\label{S:HelRev}

To set the stage, we show in \fig{Helicity_Height} vertical profiles of
$\braxy{H_{\rm M}}$, $\braxy{H_{\rm M}}/\braxy{\vec{B}^2}$, and the plasma
beta, $2\mu_0\braxy{P}/\braxy{\vec{B}^2}$ \citep[cf.][]{Bourdin:2017_beta}.
For comparison, we also plot the corresponding profiles for averages over
the AR core and the complementary quiet-Sun (QS) area.
For the plasma beta, we also show minimum and maximum values (dotted).
In the lower part, for $z\la5\Mm$, $\braxy{\vec{A}\cdot\vec{B}}$ is positive,
while for $z\ga5\Mm$ it is negative.
In fact, for the full field of view (FOV), and in a small $z$ interval
very close to the lower boundary, the sign
of $\braxy{\vec{A}\cdot\vec{B}}$ changes once again.
We return to this aspect again later.

\graph{AR_forcefree-height_VAR378}{AR_forcefree}{Force-free parameters
$\kappa_{\vec{J}\cdot\vec{B}}$ and $\kappa_{\vec{J}\times\vec{B}}$
for the AR core area and the complementary QS area.
At $z=10\Mm$, the magnetic field is nearly fully force free, so
$\kappa_{\vec{J}\times\vec{B}}\to0$ and $\kappa_{\vec{J}\cdot\vec{B}}\to1$.}

\graphside{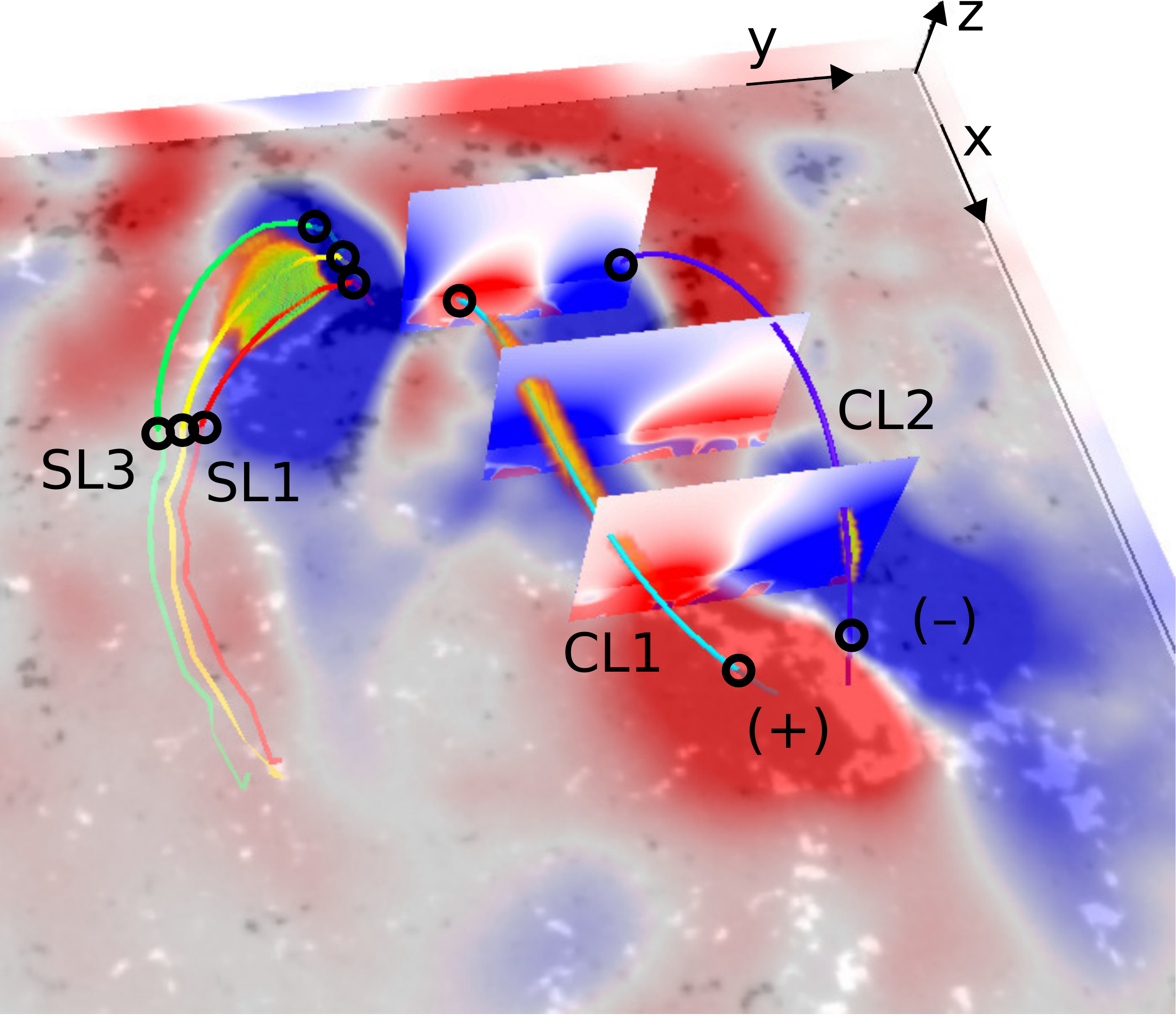}{hel.visual}{Visualization of
coronal loops (field lines) with EUV emission (orange--green volume
rendering) above an AR magnetogram (grayscale at the bottom).
The semi-transparent layer at $5.5\unit{Mm}$ shows the magnetic helicity
density with positive to negative values color-coded from red to blue, saturated
at $\pm0.2 \cdot 10^{-3}\unit{T^2 Mm}$.
The black circles mark where the field lines cross this horizontal layer.
Three vertical opaque planes cut through the cross section of the loops
in the core of the AR and also indicate the magnetic helicity density,
but saturated at $\pm0.1 \cdot 10^{-3}\unit{T^2 Mm}$;
see \sect{Results:reversal}.}

\graphwidth{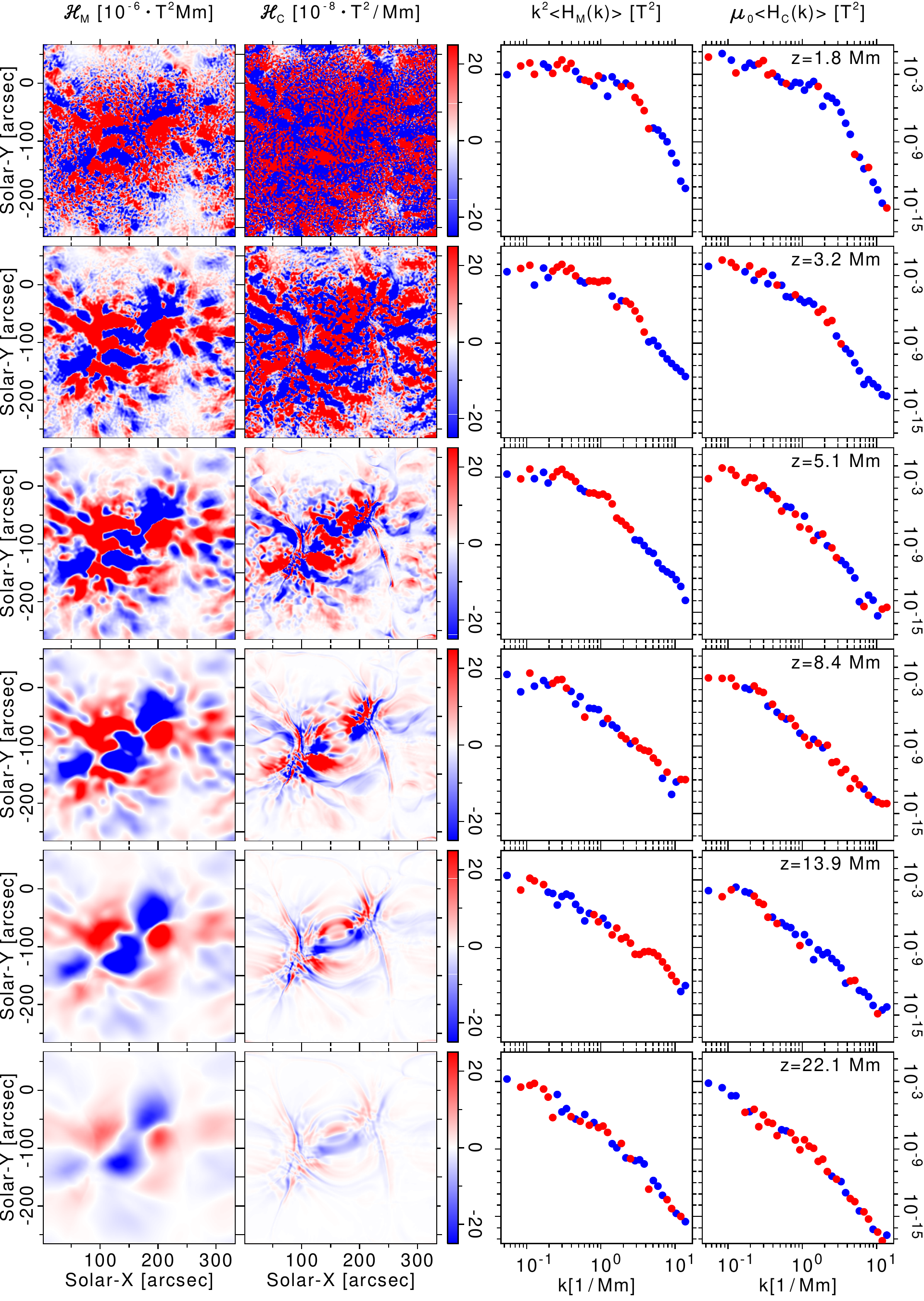}{Helicity_Maps}{Horizontal maps of the
magnetic helicity (column~1; leftmost) and current helicity (column~2) densities
for six different heights. Spectra of the magnetic helicity (column~3)
and current helicity (column~4) densities.
Positive (negative) values are shown in red (blue).}

The main focus of this paper is instead the sign reversal of
$\braxy{\vec{A}\cdot\vec{B}}$ at $z\approx5\Mm$.
This happens at a height where the magnetic field begins to become
almost force free; see \fig{Helicity_Height}(a).
To assess this quantitatively, we plot in \fig{AR_forcefree}
vertical profiles of the characteristic nondimensional wavenumbers
$\kappa_{\vec{J}\cdot\vec{B}}$ and $\kappa_{\vec{J}\times\vec{B}}$
defined through \citep{Warnecke+Brandenburg:2010}
\eql{\kappa_{\vec{J}\cdot\vec{B}}^2\equiv
\frac{\braxy{(\vec{J}\cdot\vec{B})^2}}{\braxy{\vec{J}^2\vec{B}^2}},\quad
\kappa_{\vec{J}\times\vec{B}}^2\equiv
\frac{\braxy{(\vec{J}\times\vec{B})^2}}{\braxy{\vec{J}^2\vec{B}^2}}.\quad}
Note that $\kappa_{\vec{J}\cdot\vec{B}}^2
+\kappa_{\vec{J}\times\vec{B}}^2=1$, so the two are complementary in
the sense that when $\kappa_{\vec{J}\times\vec{B}}\to0$, we have
$\kappa_{\vec{J}\cdot\vec{B}}\to1$, and vice versa.
Looking at \fig{AR_forcefree}, we see that in $5\Mm\la z\la50\Mm$,
the magnetic field is indeed nearly force free and
$\kappa_{\vec{J}\cdot\vec{B}}$ reaches values
close to unity, the largest possible value.
Consequently, $\kappa_{\vec{J}\times\vec{B}}$ is very small in this range
for AR and QS.

\subsection{Magnetic helicity reversal within a flux rope}
\label{S:Results:reversal}

In \fig{hel.visual}, we show a visualization of $\vec{A}\cdot\vec{B}$.
We see that the magnetic helicity density changes in a horizontal plane
at approximately $5\Mm$.
We also show the magnetic helicity density in several $yz$ planes
through two particularly prominent magnetic field lines labeled as
CL1 and CL2, where CL1 is a EUV-emissive loop in the core of the AR.
While one of the two field lines passes through regions where
$\vec{A}\cdot\vec{B}$ is positive (red) throughout,
the other field line traverses $yz$ planes in which
$\vec{A}\cdot\vec{B}$ is positive near the apex of the line
(denoted by CL2) and negative in the $xy$ plane through $5\Mm$
(denoted by CL1).
This shows that at least one magnetic helicity reversal is possible right
in the middle of a field line or loop.

The loops SL1--3 connect from one of the main polarities to the
periphery of the AR.
SL1--3 show strongly asymmetric heating and EUV emissivity.
We find that the coronal heating is particularly strong on that side,
where SL1--3 are rooted in strong negative magnetic helicity
(blue in \fig{hel.visual}).
The other end of these side loops connects to low-helicity areas and
there we also see less heating and EUV emissivity.
Note that under the assumption of horizontal isotropy leading to \eqn{HC_k2HM},
both the current helicity and magnetic helicity spectra are related.
In particular, since the spectral magnetic helicity reflects the
large-scale properties of current helicity, it corresponds to the integral
over all large-scale patches of the current helicity density.

We also note that the most strongly heated core-loop CL1 is rooted
in two strong positive helicity regions (red in \fig{hel.visual}), and at
the same time, we find a negative helicity (blue) near the loop apex.
While we do not want to claim a direct relation of magnetic helicity and
the coronal Ohmic dissipation of currents that heats our model loops, we
need to point out that local injection of helicity is a way of transporting
magnetic energy to the corona and induce currents there.
Nonetheless, this could only tell about the volumetric heating and the
EUV emissivity is of course strongly modulated by density variations.
In particular, when the density is low, the heating per particle is high.
Therefore, one would not see a clear one-to-one correlation of helicity and
the coronal heating or EUV emission.

\subsection{Spectral magnetic helicity reversal}\label{S:SpecHelRev}

The study of magnetic helicity spectra has revealed important insights
about the nature of the turbulent dynamo; see
\cite{Brandenburg+Subramanian:2005} for a review.
Owing to magnetic helicity conservation, the $\alpha$ effect in mean-field
electrodynamics \citep{Moffatt78,Krause+Rädler:1980} can only produce
positive and negative magnetic helicities to equal amounts, but at
different length scales \citep{Seehafer:1996,Ji:1999}.
This leads to a bihelical magnetic field
\citep{Blackman+Brandenburg:2003,Yousef+Brandenburg:2003} with one sign
at the scale of the energy carrying eddies (referred to as ``small
scale'') and another sign at the scale of the domain (referred to as
``large scale'').
In the solar wind, the spectrum is also found to be bihelical,
but the signs at both small and large scales are reversed
\citep{Brandenburg+al:2011}.
In the MHD model, the photospheric structures are ``small scale'' and
smear out when reaching higher atmospheric layers.
Coronal loops then define the ``large scale'' structures.
The basic question is now whether this apparent swap in sign at small
and large scales happens abruptly at one particular height and across
all scales, or gradually through an effective shift of the spectrum
in wavenumber, as perhaps suggested by the idea of an inverse cascade
behavior, where the height in the domain plays the role of time in a
decaying MHD simulation; see \cite{Christensson+al:2001} for an example
in the cosmological context.

The result is shown in \fig{Helicity_Maps} where we 
compare visualizations of both $\vec{A}\cdot\vec{B}$ and
$\vec{J}\cdot\vec{B}$ in six horizontal planes with the
corresponding spectra $k^2H_{\rm M}(k)$ and $H_{\rm C}(k)$
obtained in the same six planes.
Note first of all that the two spectra look similar in shape and
magnitude at corresponding heights, suggesting that the relation
between them, as given in \eqn{HC_k2HM} for isotropic turbulence,
is reasonably well obeyed. The spectra vary over more than 10 orders
of magnitude, falling steeply with wavenumber, with its largest
values corresponding to the smallest few wavenumbers that dominate
the overall sign of the total integrated magnetic and current helicities.
In the first three slices up to $z\approx5\Mm$, the dominant signs
of $k^2H_{\rm M}(k)$ and $H_{\rm C}(k)$ are negative
for $k>3\Mm^{-1}$ and positive for $k<3\Mm^{-1}$.
Above this layer, the sign of $k^2H_{\rm M}(k)$ reverses abruptly
in the sense that it is now negative (positive) for $k$ smaller
(larger) than $3\Mm^{-1}$. However, the sign of $H_{\rm C}(k)$ varies
more gradually with height, showing a similar reversal only at
$z\approx13\Mm$; see also \fig{current_helicity}.
Interestingly enough, at a fixed height, both below and above the
transition layer at $z\approx5\Mm$ where the sign reversal of the
magnetic helicity occurs, the spectrum $k^2H_{\rm M}(k)$ changes its sign
in $k$ space at roughly the same value of $k$, namely at $k\approx3\Mm^{-1}$.
This supports the notion that this phenomenon is related to
a change in the relative dominance of structures of opposite sign of
$\vec{J}\cdot\vec{B}$, as discussed above in \sect{Results:reversal},
and is not due to a shift in $k$, which would be more
reminiscent of an inverse cascade-type behavior.

We reiterate that $H_{\rm M}(k)$ is gauge independent.
It is therefore important to emphasize that the magnetic and
current helicity reversals are also seen in specific wavenumber
intervals (e.g., for $k$ larger or smaller than $3\Mm^{-1}$).
Moreover, the reversals occur at the same height as those in
$\vec{A}\cdot\vec{B}$.
This supports the notion that the sign change in $\vec{A}\cdot\vec{B}$
is not compromised by its gauge dependence; see also \fig{Gauge_Dependence} and \sect{gauge.dependence}.
Furthermore, the upper boundary condition \eqn{potBC} always tends to relax $\vec{A}$ back
to zero, as any contrasts get smeared out by the potential-field extrapolation.
The $A_x$ and $A_y$ components are set through the lower boundary condition to
match the observed $B_z$ component.
Hence, any drift in $\vec{A}$ will be suppressed.

\graph{AR_jB-angle-height_VAR378}{jB_angle}{Average angle between
$\vec{J}$ and $\vec{B}$.}

\subsection{Nearly perfectly field-aligned currents}

Within the lower corona, in the range $5\Mm\la z\la50\Mm$, the
plasma beta is around $10^{-2}$ or less; see \fig{Helicity_Height}(c).
The magnetic field here is nearly fully force free; see \fig{AR_forcefree}.
In this range, the angle
\eql{\sphericalangle(\vec{J},\vec{B})=\arccos\,(\vec{J}\cdot\vec{B}/\,(|\vec{J}|\,|\vec{B}|\,))}
between $\vec{J}$ and $\vec{B}$ is on average very small.
Closer to the surface, for $z<5\Mm$, larger values can be found,
but even then the angles are hardly much larger than $\pm2^\circ$;
see \fig{jB_angle}.
Only above the AR can larger angles of up to $\pm8^\circ$ be found.

\graph{AR_k-jB-height_VAR378}{current_helicity}{Current helicity for AR and QS versus height.}

\subsection{An additional current helicity reversal}

Very near the surface, we have seen in \fig{Helicity_Height}
for the full FOV an additional
reversal in magnetic helicity very close to the surface.
Looking at a similar plot of current helicity, we see that this
secondary reversal is now more pronounced and includes even the AR.
In current helicity, the secondary reversal is seen at $z\approx2\Mm$.
Furthermore, the primary reversal occurs higher up at about $13\Mm$.
The reason for this secondary reversal becomes more plausible when
looking at the horizontal distribution of $\vec{J}\cdot\vec{B}$
in \fig{current_helicity}, which shows that there are {\em always}
nearly equally many and nearly equally large patches of both helicities.
Thus, the dominance of one sign over the other depends on small
changes in the relative strengths of structures with positive and
negative contributions to $\vec{J}\cdot\vec{B}$.
The second reversal in current helicity is obviously a real phenomenon in
the present simulations, but it is unclear whether it is also a generic
phenomenon of stratified and magnetized atmospheres in general.
Furthermore, in magnetic helicity, it was only seen in the full FOV
and not above the AR.
Comparing the maps of magnetic and current helicities shown in
\fig{Helicity_Maps}, we see that ${\cal H}_C$ is more noisy, and
therefore the additional reversal does not appear to be a systematic feature.
More important to note is that the magnetic helicity associated with
the AR is positive near the surface, exactly as would be
expected for the southern hemisphere based on an $\alpha$ effect-driven
turbulent dynamo.

Looking once more at \fig{Helicity_Maps}, it becomes clear that the
negative sign of magnetic helicity in the uppermost layers can be
associated with a single structure that persists in all the horizontal
maps of $\vec{A}\cdot\vec{B}$ between $3$ and $22\Mm$.
This persistent helicity patch is located in the AR core near the legs of
the loop CL1 that are indicated by black circles in \fig{Helicity_Maps}.
Structures of opposite sign tend to be associated with the periphery of
the core of the AR.

\graph{AR_helicity-spectra_correlation_VAR378}{helicity_correlation}{Correlation (blue) and anticorrelation (red) in the magnetic and current helicity. The sampling height $z$ and the percentage of correlated points $r$ are given in each panel.}

\subsection{Isotropy assumption for magnetic helicity spectra}

As discussed above, under the assumption of isotropy, the magnetic and
current helicity spectra are related to each other through \eqn{HC_k2HM}.
It was already clear from \fig{Helicity_Maps} that this assumption holds
reasonably well.
The purpose of this section is to analyze this in more detail.
Therefore, we show in \fig{helicity_correlation} scatter plots of
$\mu_0 H_{\rm C}(k)$ versus $k^2H_{\rm M}(k)$ for the same six height as
in \fig{Helicity_Maps}.
It turns out that most of the data points lie on the diagonal,
which covers about eight orders of magnitude.
Some of the data points, however, have mutually opposite signs,
which correspond to an anticorrelation.
Thus, \eqn{HC_k2HM} holds primarily for the moduli of $H_{\rm C}(k)$
and $H_{\rm M}(k)$.

The fact that some of the data points have the opposite sign was
already evident when examining the colors in \fig{Helicity_Maps}.
Below $z=5\Mm$, about 80\% of the points have the expected sign,
but at higher levels, the number of exceptions increases.
For large values of $|H_{\rm M}(k)|$, and especially for $z>5\Mm$,
there is a noticeable number of data points below the diagonal,
i.e., $|H_{\rm C}(k)|$ is somewhat smaller than expected.

It is clear from \fig{Helicity_Maps} that $H_{\rm M}(k)$ shows fewer
sign reversals with $k$ than $H_{\rm C}(k)$ and shows a more systematic
behavior in that sense.
One would therefore be tempted to trust the magnetic helicity spectra
more than the current helicity spectra.
However, two other considerations come to mind.
First, both spectra are intrinsically noisy and one can expect
meaningful results only after some degree of averaging.
This could be accomplished by averaging the spectra over broader
wavenumber bins.
Second, \eqn{HC_k2HM} is only valid under the assumption of isotropy.
Again, this statement only applies in the statistical sense, i.e.,
after sufficient averaging.
This is particularly evident in the present case where there is only
one AR with its resulting coronal structure.
In view of these caveats, one must say that the agreement found in
\fig{helicity_correlation} is actually rather remarkable.

\section{Conclusions} \label{S:helicity.conclusions}

The present work has elucidated the phenomenon of a magnetic
helicity reversal above a magnetized layer in general and along a coronal
loop in particular.
We have seen that this reversal is the result of a change in the
relative dominance of structures of opposite magnetic helicity.
As a consequence, in a particular simulation, this change in sign
happens abruptly.
It also happens at all wavenumbers at the same height.
Of course, given that this change of sign depends on the subtle dominance
of structures of one sign over the other, we should expect that in other
simulations or at other times in the same simulation, such a reversal
can occur at different heights.
However, we also have found that the magnetic helicity reversal happens
near the location where the plasma beta changes from values above unity to
values below unity, i.e., when the field becomes almost force free;
see the horizontal gray dashed line in \fig{Helicity_Height},
as well as the crossing red and black lines in \fig{AR_forcefree}.
This gives us for the first time a fairly strong handle on this remarkable
phenomenon of a magnetic helicity reversal above a dynamo-active region.

It is important to note that the helicity in the lower atmosphere of our simulations
has the sign expected for the southern hemisphere, even though there is
neither a direct injection of helicity nor a mechanism to break the
north--south symmetry in the model, except for the imposed photospheric
magnetogram. A possible explanation is that the dynamo and the differential
rotation inside the Sun leave imprints in the photospheric magnetic fields.
These should then be sufficient to infer the signs of the average helicities
in the lower and upper corona.
As shown in \cite{Bourdin+Brandenburg:2018_multipole}, any arrangement of
more than two spots of unequal strength implies a non-mirrorsymmetric
pattern, which can give rise to a certain sign of magnetic helicity in
the force-free magnetic field above the surface.

Thinking now about the Sun and the solar wind, we expect the magnetic
helicity reversal to occur well within the solar corona and not between
the corona and the location of the Earth.
Thus, we expect that the magnetometers on the {\em Parker Solar Probe} and
{\em SolarOrbiter} will measure the same sign of magnetic helicity as
what is observed in Earth's neighborhood, which is opposite to what
is found at the solar surface.
The sign of course should flip if one of the measurement points is magnetically
connected to the other magnetic hemisphere of the Sun, which typically happens
if one crosses the heliospheric current sheet (HCS).
This becomes more likely during high solar activity because then the HCS may
strongly deviate from the ecliptic plane.
Perhaps the only feasible way to verify a magnetic helicity reversal so close
to the surface is by determining the wavelength at which Faraday depolarization
from intrinsic coronal emission is minimized \citep{Brandenburg+al:2017}.
This would require observations at infrared and millimeter wavelengths
just above the limb.


\acknowledgments

This work is financially supported by the Austrian Space Applications Programme at the Austrian Research Promotion Agency, FFG ASAP-12 SOPHIE under contract 853994.
The results of this research have been achieved using the PRACE Research Infrastructure resource \emph{Curie} based in France at TGCC, as well as \emph{JuRoPA} hosted by the J{\"u}lich Supercomputing Centre in Germany.
Hinode is a Japanese mission developed, launched, and operated by ISAS/JAXA, in partnership with NAOJ, NASA, and STFC (UK). Additional operational support is provided by ESA and NSC (Norway).
This research was supported in part by the NSF Astronomy and Astrophysics
Grants Program (grant 1615100), and the University of Colorado through
its support of the George Ellery Hale visiting faculty appointment.

\bibliography{Literatur-PAB,literature-NS,dynamo}
\bibliographystyle{aasjournal}

\end{document}